\documentclass[12pt]{article}
\usepackage{fullpage}

\usepackage{amsthm}
\usepackage{amsfonts}
\usepackage{amsmath}
\usepackage{graphicx}
\usepackage{enumerate}
\usepackage{setspace}
\usepackage{wasysym}
\usepackage{gastex}
\usepackage[usenames]{color}
\usepackage{soul}
\usepackage{amssymb}
\usepackage{cite}
\usepackage{verbatim}
\usepackage{authblk}

\newcommand{\bea}{\begin{eqnarray}}
\newcommand{\eea}{\end{eqnarray}}

\newtheorem{theorem}{Theorem}
\newtheorem{lemma}[theorem]{Lemma}

\newtheorem{conjecture}[theorem]{Conjecture}

\newcommand{\cf}{\mathcal{C}}

\newcommand{\RED}[1]{#1}

\begin{document}
	
\title{A note on hitting maximum and maximal cliques with a stable set}

\author{Demetres Christofides\thanks{Supported by (FP7/2007-2013)/ERC grant agreement no.~259385.}}
\affil{School of Mathematical Sciences, Queen Mary University of London, London.}
%\author{Katherine Edwards\thanks{Supported by an NSERC PGS Doctoral Fellowship and a Gordon Wu Scholarship.}}
%\affil{Department of Computer Science\\Princeton University, Princeton.}
%\author{Andrew D.~King\thanks{Supported by an NSERC Postdoctoral Fellowship, an Ebco/Eppich Fellowship, and an NSERC Discovery Grant.}}
%\affil{Department of Mathematics\\Simon Fraser University, Burnaby.}
\author{Katherine Edwards\thanks{Supported by an NSERC PGS Fellowship and a Gordon Wu Scholarship.}}
\affil{Department of Computer Science\\Princeton University, Princeton, NJ}
\author{Andrew D.~King\thanks{Corresponding author.  Email: adk7@sfu.ca.  Supported by an EBCO/Ebbich Postdoctoral Scholarship and the NSERC Discovery Grants of Pavol Hell and Bojan Mohar.}}
\affil{Departments of Mathematics and Computing Science\\Simon Fraser University, Burnaby, BC}

\maketitle
\def\thepage {} % Kill pagenumbering
\thispagestyle{empty}
\pagenumbering{arabic}

% \begin{abstract}
% 	abstract
% \end{abstract}

\begin{abstract}
It was recently proved that any graph satisfying $\omega > \frac 23(\Delta+1)$ contains a stable set hitting every maximum clique.  In this note we prove that the same is true for graphs satisfying $\omega \geq \frac 23(\Delta+1)$ unless the graph is the strong product of an \RED{odd hole and $K_{\omega/2}$}.  We also provide a counterexample to a recent conjecture on the existence of a stable set hitting every sufficiently large maximal clique.
\end{abstract}

%\newpage
\section{Introduction}

Given two graphs $G$ and $H$, the \emph{strong product} of $G$ and $H$, denoted by $G\boxtimes H$, is the graph obtained by substituting each vertex in $G$ with a copy of $H$.  The graph $C_5 \boxtimes K_3$ (see Figure \ref{fig:borodin}) has appeared as an exemplary graph in several situations, including as a counterexample to Haj\'{o}s' conjecture \cite{catlin79} and as proof of tightness of the Borodin-Kostochka conjecture \cite{borodink77}, Reed's $\omega$, $\Delta$, $\chi$ conjecture \cite{reed98}, and most recently a result on hitting all maximum cliques with a stable set:

\begin{figure}
\begin{center}
\includegraphics[scale=.55]{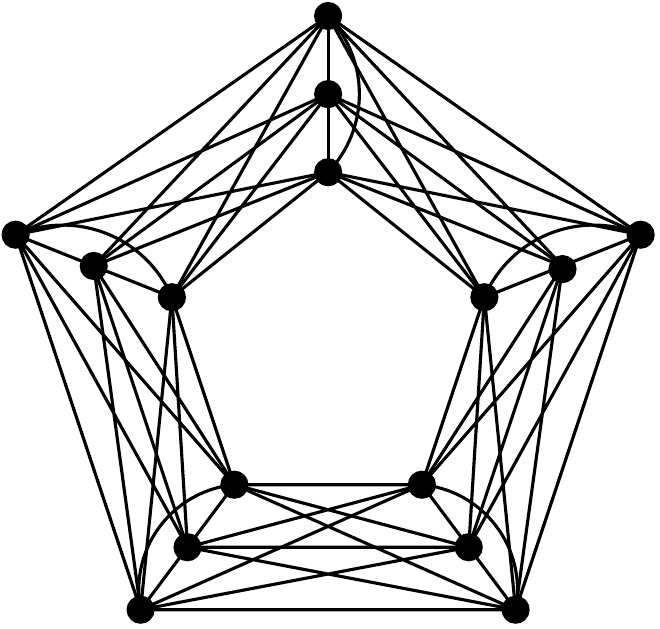}
\caption{$C_5 \boxtimes K_3$}\label{fig:borodin}
\end{center}
\end{figure}

\begin{theorem}[King \cite{king11}] \label{thm:king}
	Any graph satisfying $\omega > \frac{2}{3}(\Delta + 1)$ contains a stable set that intersects every maximum clique.
\end{theorem}

\RED{This theorem is a refinement of a result of Rabern \cite{rabern10}, who proved the result when $\omega \geq \frac 34(\Delta+1)$.  The refinement relies on a strengthening of Haxell's Theorem \cite{haxell95}; this strengthening was implicit in Haxell's work and also in work of Aharoni, Berger, and Ziv \cite{aharonibz07}.}

Since $C_5 \boxtimes K_3$ satisfies $\omega = \frac 23(\Delta+1)$ but contains no stable set hitting every maximum clique, the strict inequality in Theorem \ref{thm:king} is necessary.  Actually $C_5$ itself also shows that strictness is necessary, and is not just a Brooks-type exception.  In the next two sections of this note we prove that any graph that exhibits this property is the strong product of an odd hole\footnote{A {\em hole} is an induced cycle of length at least 4.} and a clique:

\begin{theorem} \label{thm-main}
Any connected graph satisfying $\omega \geq \frac{2}{3}(\Delta + 1)$ contains a stable set intersecting every maximum clique unless it is the strong product of \RED{an odd hole and a clique.}
\end{theorem}

It is easy to confirm that the strong product of a \RED{an odd hole and a clique} does not contain a stable set hitting every maximum clique.  In the last section of this note, we prove that there is no hope of proving a statement analogous to Theorem \ref{thm:king} for maximal rather than maximum cliques.

%%%%%%%%%%%%%%%%%%%%%%%%%%%%%%%%%%%%%%%%%%%%%%%%%%%%%%%%%%%%%%%%%%%%%%%%
%%%%%%%%%%%%%%%%%%%%%%%%%%%%%%%%%%%%%%%%%%%%%%%%%%%%%%%%%%%%%%%%%%%%%%%%
%%%%%%%%%%%%%%%%%%%%%%%%%%%%%%%%%%%%%%%%%%%%%%%%%%%%%%%%%%%%%%%%%%%%%%%%
%%%%%%%%%%%%%%%%%%%%%%%%%%%%%%%%%%%%%%%%%%%%%%%%%%%%%%%%%%%%%%%%%%%%%%%%
\section{The clique graph}

Following \cite{king11} and \cite{rabern10}, we approach Theorem \ref{thm-main} by characterizing the structure of the {\em clique graph}.  Given a graph $G$ and a collection $\mathcal{C}$ of maximum cliques in $G$, we define the clique graph, denoted by $G(\mathcal{C})$, as follows. The vertices of $G(\mathcal{C})$ correspond to the cliques in $\mathcal{C}$; two vertices of $G(\mathcal C)$ are adjacent if and only if their corresponding cliques intersect in $G$.%\footnote{Some may find it interesting to note that if $G$ is triangle-free, then the clique graph of $G$ is isomorphic to the line graph of $G$.}.

For now we can restrict our attention to connected clique graphs.  When $\omega > \frac 23(\Delta+1)$, we are guaranteed that if $G(\mathcal C)$ is connected, then $|\cap \mathcal C|\geq \frac 13(\Delta+1)$ \cite{king11}.  However, the same is not necessarily true when $\omega = \frac 23(\Delta+1)$, for example with the strong product of either \RED{a hole (i.e.\ a cycle of length $\geq 4$) and a clique, or $P_\ell$ (i.e.\ a path on $\ell$ vertices) for $\ell \geq 4$ and a clique}, in which case $\cap \mathcal C$ is empty.  This is actually the only troublesome case.  To prove this we need Hajnal's set collection lemma.

\begin{lemma}[Hajnal \cite{hajnal65}] \label{lem-hajnal}
	Let $G$ be a graph and let $\mathcal C$ be a collection of maximum cliques in $G$. Then \[|\cap\mathcal C| + |\cup\mathcal C| \geq 2\omega(G).\]
\end{lemma}

The following lemma extends a lemma of Kostochka \cite{kostochka80} that is instrumental to the proof of Theorem \ref{thm:king}.

\begin{lemma}\label{lem:main}
Suppose $G$ is connected and satisfies $\omega \geq \frac 23(\Delta+1)$, and let $\mathcal C$ be a collection of maximum cliques in $G$ such that $G(\mathcal C)$ is connected \RED{and $|\cap \mathcal C | < \frac 13(\Delta+1)$.  Then $\cap \mathcal C = \emptyset$, and for some $k\geq 4$ either $G$ is $C_k\boxtimes K_{\omega/2}$, or the subgraph induced by $\cup\mathcal C$ contains $P_k\boxtimes K_{\omega/2}$ as a subgraph.}
\end{lemma}

\RED{Kostochka's lemma (which appears in English in \cite{king11} and \cite{rabern10}) actually tells us that if $\omega > \frac 23(\Delta+1)$, no such set $\mathcal C$ can exist.  So it suffices to deal with the case $\omega = \frac 23(\Delta+1)$.}

\begin{proof}
\RED{Assume $\omega = \frac 23(\Delta+1)$.  Note that if $\mathcal C'$ is any family of maximum cliques with $\cap \mathcal C'\neq \emptyset$, then $|\cup \mathcal C'|\leq \Delta+1$.  Otherwise, every vertex in $\cap \mathcal C'$ would have more than $\Delta$ neighbours, which is impossible.

For any two intersecting maximum cliques $A$ and $B$, we know by the previous paragraph that $|A\cap B| = 2\omega - |A\cup B| \geq 2\omega - (\Delta+1) = \omega/2$.  Now let $\mathcal C'$ be a maximal set of cliques such that $|\cap \mathcal C'|\geq \omega/2$, and let $A$ and $B$ be two intersecting cliques in $\mathcal C'$ such that $B$ intersects a clique $C$ in $\mathcal C\setminus \mathcal C'$ (we know that $|\mathcal C|\geq 3$ because $|A\cap B|\geq \omega/2$, so this must be possible since $G(\mathcal C)$ is connected).  Let $\mathcal C''$ denote $\mathcal C'\cup \{C\}$.

By the maximality of $\mathcal C'$, we have $|\cap \mathcal C''| < \frac 13(\Delta+1)$.  Suppose that $\cap \mathcal C''$ is nonempty.  Any vertex in $\cap{\mathcal C''}$ is adjacent to the rest of $\cup{\mathcal C''}$, so $|\cup\mathcal C''| \leq \Delta+1$.  But this contradicts Lemma \ref{lem-hajnal}, so $\cap{\mathcal C''}$ must indeed be empty and therefore $\cap \mathcal C = \emptyset$.

%Let $\mathcal{C}$ be a collection of maximum cliques such that $|\cap \cf| < \frac{1}{3}(\Delta + 1)$ and such that $G(\mathcal C)$ is connected.  Now let $\mathcal{C}' \subset \mathcal{C}$ be a maximal collection of cliques with ${|\cap\mathcal{C}'| \geq \frac{1}{3}(\Delta + 1)}$.  Since a vertex in $\cap\mathcal C$ is adjacent to every other vertex in $\cup\mathcal{C'}$, $|\cup\mathcal{C'}|$ must be at most $\Delta+1$.  And since $|\cup \mathcal C'| = 2\omega - |\cap\mathcal C'| \leq \Delta+1$, any pair of intersecting cliques in $\mathcal C$ intersect in at least $\frac 13(\Delta+1)$ vertices.  Thus $|\mathcal C'|\geq 2$.  

%  Now let $C_1$ and $C_2$ be two intersecting cliques in $\mathcal C$ such that $C_1 \in \mathcal C'$ and $C_2\notin \mathcal C'$.  Let $\mathcal C''$ denote $\mathcal C'\cup \{C_2\}$.  By the maximality of $\mathcal C'$, we have $|\cap{\mathcal C''}|<\frac 13(\Delta+1)$.  Suppose that $\cap{\mathcal C''}$ is nonempty.  Any vertex in $\cap{\mathcal C''}$ is adjacent to the rest of $\cup{\mathcal C''}$, so $|\cup\mathcal C''| \leq \Delta+1$.  But this contradicts Lemma \ref{lem-hajnal}, so $\cap{\mathcal C''}$ must indeed be empty and therefore $\cap \mathcal C = \emptyset$.

Since $B\cap C \neq \emptyset$ it follows that $|B\cap C| \geq \omega/2$. On the other hand we also have $|B \setminus C| \geq |\cap \mathcal{C}'| \geq \omega/2$ and so $|B \cap C| = |\cap \mathcal{C}'| = \omega/2$.  Thus it is clear that the sets $(B \cap C)$ and $(\cap \mathcal C')$ partition $B$.  Also, no clique of $\mathcal C'$ can intersect $C\setminus B$, since a vertex in this intersection would be complete to $B$, contradicting the fact that $B$ is a maximum clique.  Further, no clique $D$ of $\mathcal C'$ other than $B$ can intersect $C$, since this would imply that $D$ and $C$ have nonempty intersection of size less than $\omega/2$, which is impossible.  Therefore $\mathcal C' = \{A,B\}$, otherwise $|\cup \mathcal C'|$ would be greater than $\Delta+1$.

We have shown that $|\mathcal C|\geq 3$, and given any three cliques $A,B,C\in \mathcal C$ with $|A\cap B\cap C|< \omega/2$ such that $A$ and $C$ both intersect $B$,
\begin{enumerate}
\item $A$ and $C$ are disjoint,
\item $A\cap B$ and $C\cap B$ have size $\omega/2$ and partition $B$, and
\item no other maximum clique $D$ intersects $B$.
\end{enumerate}
It follows that $G(\mathcal C)$ has maximum degree 2 (and by assumption, is connected).  Therefore the subgraph induced by $\cup \mathcal C$ contains, for some $k\geq 4$, either $P_k\boxtimes K_{\omega/2}$ or $C_k\boxtimes K_{\omega/2}$ as a subgraph.  Finally, since $G$ is connected and $C_k\boxtimes K_{\omega/2}$ is $(\frac 32\omega-1)$-regular, if $G$ contains $C_k\boxtimes K_{\omega/2}$ as a subgraph then $G$ is isomorphic to $C_k\boxtimes K_{\omega/2}$.  This completes the proof.}
\end{proof}

%%%%%%%%%%%%%%%%%%%%%%%%%
%%%%%%%%%%%%%%%%%%%%%%%%%
%%%%%%%%%%%%%%%%%%%%%%%%%
%%%%%%%%%%%%%%%%%%%%%%%%%
\section{Hitting the maximum cliques with a stable set}
%%%%%%%%%%%%%%%%%%%%%%%%%

In order to find our desired stable set, we need the main intermediate result in the proof of Theorem \ref{thm:king}, \RED{which extends Haxell's Theorem \cite{haxell95}.}

\begin{theorem}[King \cite{king11}] \label{thm-isr}
\RED{Let $G$ be a graph with vertices partitioned into cliques $V_1,\dots, V_r$, and let $k$ be a positive integer.} If for every $i$ and every $v \in V_i$, $v$ has at most $\min\{k, |V_i|-k\}$ neighbours outside $V_i$, then $G$ contains a stable set of size $r$.
\end{theorem}

\begin{proof}[Proof of Theorem \ref{thm-main}]

For fixed $\omega(G)\geq 1$ we proceed by induction on $|V(G)|$; the result trivially holds whenever $|V(G)|\leq \omega(G)$.  Let $\cf$ be the set of maximum cliques in a graph $G$, and let $\cf_1, \cf_2,\ldots, \cf_k$ be the partitioning of $\cf$ such that $G[\cf_1],G[\cf_2],\ldots,G[\cf_k]$ are the connected components of the clique graph $G[\cf]$.  We consider two cases.  The first case is basically the same as the proof of Theorem \ref{thm:king}.\\

\noindent{\bf Case 1:} For every $1\leq i \leq k$, $\cap \cf_i \neq \emptyset$.

By Lemma \ref{lem:main}, for every $1\leq i \leq k$ we have $|\cap \cf_i|\geq \frac 13(\Delta(G)+1)$.  It suffices to show that there is a stable set in $G$ intersecting each $\cap \cf_i$.  For a given $i$, every vertex in $\cap \cf_i$ has at most $\Delta(G)+1 -|\cup \cf_i|$ neighbours in $\cup_{j\neq i}(\cap \cf_j)$.  Lemma \ref{lem-hajnal} tells us that $|\cup \cf_i|+|\cap \cf_i| \geq \frac 43(\Delta(G)+1)$.  Therefore $\Delta(G)+1-|\cup \cf_i|\leq |\cap \cf_i|-\frac 13(\Delta(G)+1)$.  And since $|\cup\cf_i|\geq \omega(G) \geq \frac 23(\Delta(G)+1)$, a vertex in $\cap\cf_i$ has at most $\min\{ \frac 13(\Delta(G)+1), |\cap\cf_i|-\frac 13(\Delta(G)+1)  \}$ neighbours in $\cup_{j\neq i}(\cap \cf_j)$.  It therefore follows from Theorem \ref{thm-isr} that there is a stable set in $G$ intersecting each $\cap \cf_i$.  This completes Case 1.\\

\noindent{\bf Case 2:} For some $1\leq i\leq k$, $\cap \cf_i=\emptyset$.

Assume that $\cap\cf_1=\emptyset$.  Lemma \ref{lem:main} tells us that either $G$ is the strong product \RED{a hole and $K_{\omega(G)/2}$}, or $G[\cup \cf_i]$ contains as a subgraph the strong product of $K_{\omega(G)/2}$ and a $P_\ell$ for $\ell\geq 4$.  In the former case the theorem clearly holds, so let us consider the latter case.  If there is a vertex not in a clique of size $\omega(G)$, we can delete it and apply induction, so assume that no such vertex exists.  Let the cliques of $\cf_1$ be $C_1,\ldots,C_{\ell-1}$ such that for $1\leq i \leq \ell-2$, $C_i$ and $C_{i+1}$ intersect in exactly $\omega(G)/2$ vertices.  Let $X_1$ denote $C_1\setminus C_2$ and let $X_2$ denote $C_{\ell-1}\setminus C_{\ell-2}$.

\begin{figure}
\begin{center}
\includegraphics[scale=.7]{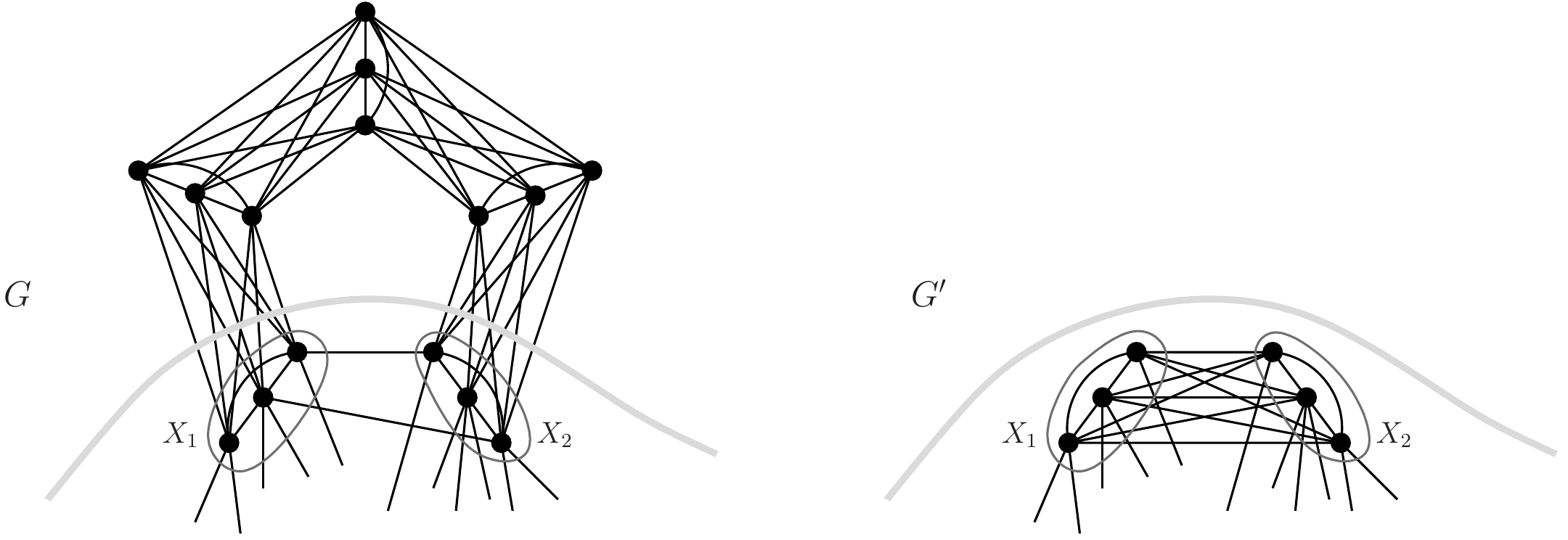}
\caption{A reduction of a clique path for $\ell = 5$}\label{fig:reduction}
\end{center}
\end{figure}

We will construct a graph $G'$ on fewer than $|V(G)|$ vertices such that $\omega(G')=\omega(G)$ and $\Delta(G')\leq\Delta(G)$, and apply induction to prove our result.  To construct $G'$ from $G$ we delete $\cup_{1\leq i \leq \ell-2}(C_i\cap C_{i+1}) = (\cup \cf_1 )\setminus(X_1\cup X_2)$ and add edges to make $X_1\cup X_2$ a clique of size $\omega$ in $G'$ (see Figure \ref{fig:reduction}).  Clearly $G'$ has maximum degree at most $\Delta(G)$.  We claim that $G'$ has clique number $\omega(G)$.  Suppose this is not the case.  It follows that there exists a set $Y_1\subseteq X_1\cup X_2$ and a set $Y_2$ in $V(G)\setminus \cup \cf_1$ such that $Y_1\cup Y_2$ is a clique of size greater than $\omega(G)$.  Let $v$ be a vertex in $Y_2$.  Since $v$ is in an $\omega(G)$-clique in \RED{$G\setminus (X_1\cup X_2)$}, it has at most $\omega(G)/2$ neighbours in $X_1\cup X_2$, so $|Y_1|\leq \omega(G)/2$.  Therefore $|Y_2|>\omega(G)/2$, which implies that some vertex in $Y_1$ has at least $\omega(G)-1$ neighbours in $\cup\cf_1$ and more than $\omega(G)/2$ neighbours in $Y_2$, contradicting the fact that $\omega(G) \geq \frac 23(\Delta(G)+1)$.  Therefore $G'$ has clique number $\omega(G)$.

By induction, there is a stable set $S$ in $G'$ hitting every $\omega(G)$-clique.  Thus $S$ is also a stable set in $G$ intersecting $X_1\cup X_2$ exactly once.  Without loss of generality let $v$ be a vertex in $X_1\cap S$.  From $S$ we will construct a stable set $S'$ hitting every $\omega(G)$-clique in $G$ in one of two ways, depending on the parity of $\ell$.

If $\ell$ is even, let $S'$ consist of $S$ along with one vertex in $C_{2k}\cap C_{2k+1}$ for each $1 \leq k \leq (\ell/2) -1$.  It is a routine exercise to confirm that $S'$ is a stable set hitting every maximum clique in $G$.

If $\ell$ is odd, let $S'$ consist of $S\setminus \{v\}$ along with \RED{one vertex from} $C_{2k-1}\cap C_{2k}$ for each $1 \leq k \leq (\ell-1)/2$.  Again $S'$ is a stable set hitting every maximum clique in $G$, because the only $\omega(G)$-clique intersecting $C_1\setminus C_2$ is $C_1$.  This completes the proof.
\end{proof}

%%%%%%%%%%%%%%%%%%%%%%%%%
%%%%%%%%%%%%%%%%%%%%%%%%%
%%%%%%%%%%%%%%%%%%%%%%%%%
%%%%%%%%%%%%%%%%%%%%%%%%%
\section{Hitting large maximal cliques with a stable set}
%%%%%%%%%%%%%%%%%%%%%%%%%

Theorem \ref{thm:king} can be used to characterize minimum counterexamples to Reed's $\chi$, $\omega$, $\Delta$ conjecture; see for example \cite{aravindks11} \S4.  Motivated by the problem of similarly characterizing minimum counterexamples to the local strengthening of Reed's $\chi$, $\omega$, $\Delta$ conjecture (see \cite{chudnovskykps11, kingthesis}), King recently proposed the following unpublished conjecture:

\begin{conjecture}
There exists a universal constant $\epsilon > 0$ such that every graph contains a stable set hitting every maximal clique of size at least $(1-\epsilon)(\Delta+1)$.
\end{conjecture}

We conclude this note by disproving the conjecture.

\begin{theorem}
For any $\epsilon>0$ there exists a graph in which every maximal clique has size at least $(1-\epsilon)(\Delta+1)$, and no stable set hits every maximal clique.
\end{theorem}

\begin{proof}
Choose two positive integers $k$ and $t$ sufficiently large such that 
\begin{equation}\label{eq1}
(1-\epsilon)(kt +5t-5) < kt +2-k.
\end{equation}
We now construct a graph with vertices partitioned into sets $A$ and $B$ of size $kt$ and $5t$ respectively.  We further partition $A$ into $A_1,\ldots, A_t$ and $B$ into $B_1,\ldots,B_t$ such that
\begin{enumerate}
\item $A$ is a clique and each $A_i$ has size $k$
\item each $B_i$ induces a $5$-cycle, and there are no edges between $B_i$ and $B_j$ for $i\neq j$
\item vertices $u\in A_i$ and $v\in B_j$ are adjacent precisely when $i\neq j$.
\end{enumerate}
Thus we can see that the unique maximum clique in $G$ is $\cup_i A_i$, with size $kt$.  All other maximal cliques of $G$ consist of two vertices in $B$ and $k(t-1)$ vertices of $A$.  The maximum degree of the graph is $kt+5t-6$, achieved by all vertices in $A$.  By (\ref{eq1}), every maximal clique has size greater than $(1-\epsilon)(\Delta+1)$.

It therefore suffices to prove that no stable set intersects every maximal clique.  Suppose we have a stable set $S$ intersecting every maximal clique.  Since $A$ is a maximal clique, without loss of generality we can assume $S$ intersects $A_1$, and therefore $S\setminus A_1 \subseteq B_1$.  But then there must remain two adjacent vertices in $B_1\setminus S$.  Together with $\cup_{j\neq 1}A_j$ these vertices form a maximal clique in $G$.  This contradiction completes the proof.
\end{proof}

%%%%%%%%%%%%%%%%%%%%%%%%%
%%%%%%%%%%%%%%%%%%%%%%%%%
%%%%%%%%%%%%%%%%%%%%%%%%%
%%%%%%%%%%%%%%%%%%%%%%%%%
\section{Acknowledgements}
%%%%%%%%%%%%%%%%%%%%%%%%%

\RED{This work was done at ITI in Prague; the authors are grateful to everyone involved, particularly Daniel Kr\'al', for their hospitality.  The authors would also like to thank the two anonymous referees for their careful reading and helpful suggestions.}

\bibliography{masterbib}{}
\bibliographystyle{plain}

\end{document}